# A Typology of Collaboration Platform Users

Anastasia Bezzubtseva[1,2], Dmitry Ignatov[1]

[1]Higher School of Economics, Moscow, Russia
[2]Witology, Moscow, Russia

nstbezz@gmail.com, dignatov@hse.ru

**Abstract.** In this paper we present a review of the existing typologies of Internet service users. We zoom in on social networking services including blogs and crowdsourcing websites. Based on the results of the analysis of the considered typologies obtained by means of FCA we developed a new user typology of a certain class of Internet services, namely a collaboration innovation platform. Cluster analysis of data extracted from the collaboration platform Witology was used to divide more than 500 participants into 6 groups based on 3 activity indicators: idea generation, commenting, and evaluation (assigning marks) The obtained groups and their percentages appear to follow the "90 − 9 − 1" rule.

**Keywords.** Crowdsourcing, typology classification, collaborative platform, innovation, social network, community, blog.

## 1    Introduction

Collaboration innovation platforms are relatively young and less common than blogs or social networks (e.g., compare [1] and [2]), yet interest in their organization and audience is not decreasing. The existing studies of consumer or media behavior of Internet users cannot be fully applied to collaboration platform participants, while general psychological or sociological typologies of people miss many important features, inherent only to networking and crowdsourcing.

For a certain type of social network services, i.e. the collaboration innovation platforms, finding user types pursues also some other objectives. Understanding user types could make a major contribution in the platform effectiveness. For instance, dynamic participant type detection and displaying are useful as a motivational game component, and the type itself will probably supplement or refine the exiting rating systems. Also, information about the amount of users of different groups could help platform moderators turn community life to a beneficial for invention direction.

In this study we present a review of the existing Internet service user classifications. Based on examined materials we attempted to develop a new typology of collaboration platform participants using data of one of the projects of Russian innovation platform Witology [10].

## 2    Terminology

In this paper we analyze not only collaboration platforms, but also all other kinds of social networking services and Internet services, as typologies of their users can be applied to platform participants. There is no fixed terminology in this area yet, but we will try to give some definitions of the important concepts used in the research in order to clarify its subject.

By *Internet service* we mean any website that provides any kind of service (e.g. blogs, file-sharing networks, chats, multiplayer games, online shops). Internet services which provide human interaction are referred to as *Social Networking Services* (SNS). They include social networks (Facebook, MySpace, last.fm, LinkedIn, Orkut), blogs (LiveJournal, Tumblr, Twitter), wiki (e.g., Wikipedia), media hosting sites (Flickr, Picasa, YouTube), etc. [3], [4]. Social networking services often generate *online communities, i.e.* groups of people, who share similar interests and communicate via a certain Internet service. Some scientists [5], [6], [7] understand community in a wider sense as the entire audience of some social networking service, which is wrong, according to Michael Wu [8]. We kept the original author vocabularies when describing the typologies, in other cases the first definition of community was used.

*Crowdsourcing platforms* are social networking services which are used to obtain the necessary services, ideas or content from platform participants, i.e. platform community, as opposed to regular staff or vendors [9]. *Crowdsourcing (collaboration) innovation platforms* are the ones which focus on idea generation. Activities on collaboration platforms often include message (idea or comment) posting, message reading and message evaluation. The winning solutions and true experts are identified on the basis of the amount and quality of such activities. Work on the platform usually goes as a certain time-limited project, devoted to some company's problem. Witology [10], Imaginatik [11], BrightIdea [12] and some other platforms are organized this way; though, there are many collaboration sites which are not alike (see list [13]).

## 3    Research objectives

To begin a classification of collaboration innovation platform users, we plan to perform the following tasks:
1. *Study of the existing Internet service user typologies.* The discovered user types and data mining techniques might be helpful in developing another typology.
2. *Developing of a new typology of collaboration innovation platform.* By means of mathematical methods we plan to analyze data of one of the collaboration platform project and identify distinct user types.
3. *Comparison of the obtained percentages with the ones from existing studies.* This might help to understand whether the community under analysis is typical and to find out, whether it can be improved (for example, by calculating community health index [14]).

## 4      Review of the existing typologies

Despite the fact that the online community being a relatively young phenomenon, tens of attempts in classifying internet users have been undertaken. Some of the studies [15], [16], [17] explore only children's media-behavior, others [18] investigate behavior in terms of online shopping. A significant part of early typologies (e.g. [19], [20]) is developed based on frequency and variety of web and new gadgets use, which resulted in rather trivial and similar typologies (generally people were divided into "advanced", "average" and "non-users", the three types were occasionally interspersed with "entertainment" and "functional" users).

Almost half of the encountered researches used cluster analysis as means of extracting user types, factor analysis appeared to be the second most popular method. Much more uncommon were regression analysis, qualitative in-depth analysis, graph mining, statistical analysis, etc.

Very few authors based on some sociological or psychological theories or referred to the existing typologies when classifying internet service users (it can be explained by their desire to take a new look on the differences in human behavior). One of the studies (Nielsen, 2006) [7] is not only descriptive, but is considered informal, and in spite of that the classification and the "90 – 9 – 1" rule are highly respected and popular.

As for the user typologies of the communities, which organization is close to that of innovation platforms, a notable part of papers is devoted to social network user behavior analysis, but there are also some studies of behavior of blog and forum visitors. Since information concerning behavior of collaboration platform participants has not been found yet, several of social network and blog studies might be interesting and useful as a basis for development of an original classification of collaboration platform users. Further we describe those relevant typologies.

### 4.1      Describing user typologies

**Brandtzæg and Heim (2010).** The study [5] is a descriptive one, though the list of existing theories and research papers is given in one of its sections. The results of online survey of 4 Norway social networks users were subjected to cluster analysis.

- *Sporadics* visit social network from time to time, mainly to check if somebody contacted them.
- *Lurkers* is the largest group, they do not create any content, but consume and spread the content created by other groups. They are also notable for a propensity to time-killing.
- *Socializers* use social networks to communicate, make new friends, comment on photos of the old ones, post congratulation messages on walls etc.
- *Debaters* are a more mature and educated version of socializers. Besides communication, less shallow than in the previous case, they are interested in consumption and discussion of news and other information available in social networks.

- *Actives* are engaged with all possible types of activity: communication, reading, creating, watching, establishing groups.

**Budak, Agrawal, Abbadi (2010).** This paper [13] describes the three types of people (presented in 2002 by Malcolm Gladwell [21]) in terms of graph theory in context of modern online communities (especially blogs). The presence of those people, in Gladwell's opinion, is the main cause of the resounding popularity of some innovations. Authors also introduce a new type (the Translators), which, along with the Sellers, more than other groups influences idea spread and success.

- *Connectors* are people who easily make friends and, thus, have a lot of them.
- *Mavens* are very informed due to their curiosity and like to share their knowledge.
- *Salesmen,* – it is natural for them to convince people and establish an emotional contact with them.
- *Translators* are "bridges" between different interest groups. They have the ability to interpret ideas in a different way, so that more people could understand and accept them.

**Li, Bernoff, Fiorentino, and Glass (2007) present** another classification [25] without theoretical basis. Groups were extracted with the help of cluster analysis of the poll values.

- *Creators* blog, publish video, maintain their own web-sites; usually belong to the young generation.
- *Critics* select and choose useful media content; typically older than the previous group.
- *Collectors* are known for their addiction to saving bookmarks on special services.
- *Joiners* spend much time in social networks; the youngest group.
- *Spectators* read blogs, watch video, listen to podcasts; main consumers of user-generated content.
- *Inactives* are not active in social services.

**Nielsen (2006).** In the study [7] it is assumed that active members of large communities are very few. No special mathematical instruments were used to develop the typology, although the author mentions that user activity follows Power law (in the Zipf curve variant).

- *Lurkers* (90%) are those who only consume.
- *Intermittent/sporadic contributors* (9%) are those who contribute rarely, occasionally.
- *Heavy contributors/active participants* (1%) are responsible for up to 90% of community materials.

**Jepsen (2006).** This is one of the few classifications [23] with a theoretical foundation (Kozinetz, 1999) [22]). The members of Danish newsgroups were classified according to mean and median survey values.

- *Tourists* are not very interested in community content.
- *Minglers* are sociable people, who prefer not to consume the site's content, but to communicate with other members.
- *Devotees* are compared to minglers more interested in newsgroup materials than in communication.
- *Insiders* both communicate and consume information.

**Golder and Donath (2004).** This is one more descriptive study [24] which examined 16 unmoderated Usenet newsgroups. The taxonomy was built after in-depth analysis of the message posting frequency and message content.

- *Celebrities* are central community figures, contribute more than others.
- *Newbies* are new members, which ask many questions and do not know how to act and communicate appropriately.
- *Lurkers* are those who read discussions, but do not take part in them.
- *Flamers, Trolls, Ranters* – three subgroups, members of which are notable for their negative behavior and love to conversation spoiling.

## 4.2    Comparing user typologies

Analysis of the mentioned typologies resulted in an assumption that, despite some significant differences in social networking services, there is a universal set of user types. Though, some sources claim that there could be no such a meta-typology [23], when others [6] make attempts in developing one.

The resemblance of user types can be seen more clearly from table 1. Also some insights could be provided by a formal concept lattice, derived from the table (fig. 1). Rows of the table represent the user types described previously (objects), columns are the relevant typologies (attributes). Similar classes were merged: thus, class Actives of the table includes Actives (Brandtzaeg & Heim, 2010), Active participants (Nielsen, 2006), Insiders (Jepsen, 2006), and Celebrities (Golder & Donath, 2004).

**Table 1.** Formal context (types as objects, typologies as attributes)

|  | Brandtzaeg & Heim (2010) | Budak et al. (2010) | Li et al. (2007) | Nielsen (2006) | Jepsen (2006) | Golder & Donath (2004) |
|---|---|---|---|---|---|---|
| Inactives | 1 | 0 | 1 | 0 | 1 | 0 |
| Lurkers | 1 | 0 | 1 | 1 | 1 | 1 |
| Socializers | 1 | 1 | 1 | 0 | 1 | 0 |
| Debators | 1 | 0 | 1 | 0 | 0 | 0 |
| Actives | 1 | 0 | 0 | 1 | 1 | 1 |
| Salesmen | 0 | 1 | 0 | 0 | 0 | 0 |
| Translators | 0 | 1 | 0 | 0 | 0 | 0 |

| Collectors | 0 | 0 | 1 | 0 | 0 | 0 |
| Creators | 0 | 1 | 1 | 1 | 0 | 0 |
| Newbies | 0 | 0 | 0 | 0 | 0 | 1 |
| Negatives | 0 | 0 | 0 | 0 | 0 | 1 |

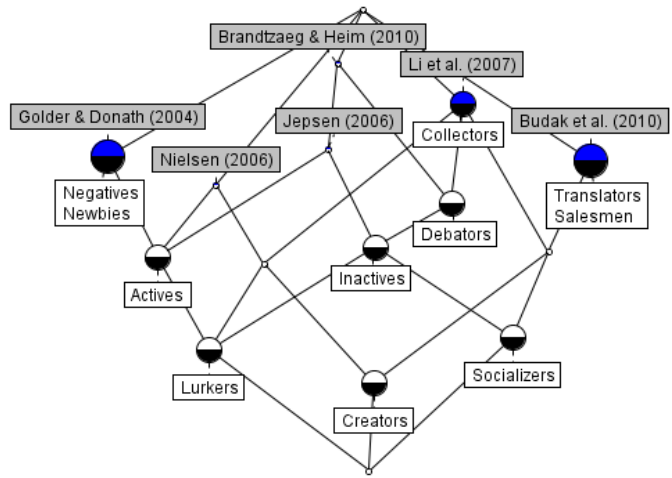

**Fig. 1.** Formal concept lattice of user typologies (built in ConExp [24])

It can be assumed from the picture that the three general classes of users at the bottom (Lurkers, Creators and Socializers) and, perhaps, two or three important, but less general classes (concepts) above (Actives, Inactives, Debators) form a universal classification of social networking service users. It can also be seen that three studies introduced five original user classes (Negatives, Newbies, Collectors, Translators, Salesmen), which are less likely to be found in a community. As for the typologies, the one of Brandtzaeg & Heim (2010) appears to be the most common.

We built Duquenne-Guigues base for the context and selected the implications with support greater than 4:

1. supp = 4, Actives ==> Lurkers;
2. supp = 3, Inactives ==> Lurkers Socializers;
3. supp = 3, Lurkers Socializers ==> Inactives;
4. supp = 2, Debators ==> Inactives Lurkers Socializers.

E.g., implication 1 can be read as "Each user typology which contains Actives also contains Lurkers and it is valid in 4 cases out of 6".

# 5 Typology construction and analysis

## 5.1 Data sample

We used data obtained in one of the projects [25] of the collaboration platform Witology. It includes quantitative indicators of each of participants' activity: the number of generated ideas, the number of posted comments and the number of submitted evaluations.There were also some other types of activities on the platform, but the mentioned ones are the most basic and easy to interpret.

The project administrators and moderators were not considered as a part of a crowdsourcing community, so only 504 of all 519 registered platform users were sampled.

## 5.2 Analysis

Initially we detected those participants, who never commented, evaluated or generated ideas. These 248 users were clearly not interested in the project (165 of them never logged on the platform after the third day of its work); thus, they could be excluded from the further analysis.

Then we used clustering algorithm (k-means [26]) to divide the sample based on several parameters. The results of cluster analysis of 256 objects are presented in fig.1 (we used XLSTAT 2011 [27] for the analysis, and XLSTAT-3DPlot package for visualization).

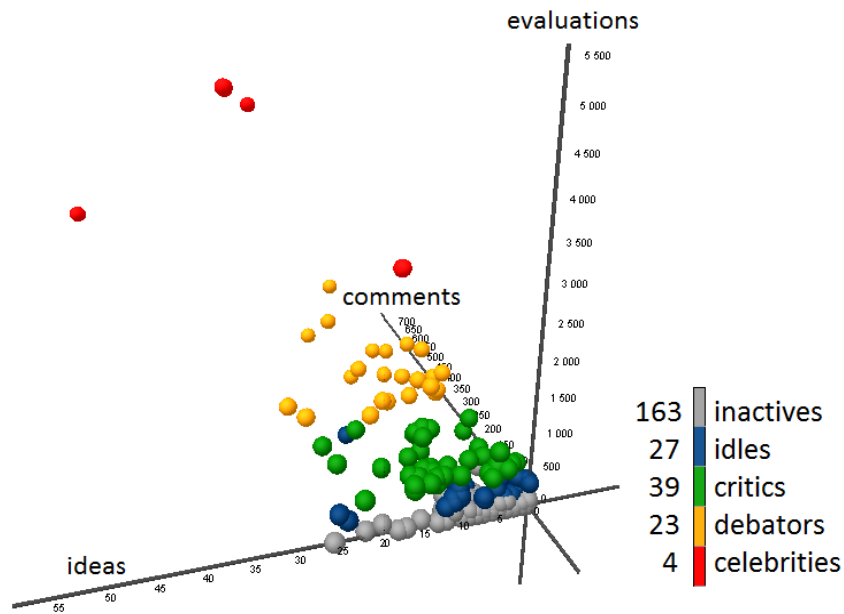

**Fig. 2.** Sample clustering (the number of objects in each cluster is displayed to the left to the color scale)

The first cluster (grey) represents the participants, who did not show much activity in evaluation and commenting. Because of the difference in orders of numbers of created ideas, comments and evaluations, the participants who seem to be prominent idea generators (created more than 10 ideas) ended up in this group.

The second cluster (blue) differs from the first with slightly higher evaluation activity of users. It can be assumed that those people were interested in project, but lacked motivation for message posting. It is reasonable to merge a certain part of this cluster with the previous one.

The third cluster (green) as a whole is hard to characterize. Its members are less passive: they may skip idea generation or comment posting, but they always evaluate something.

The fourth cluster (yellow) is not far from the previous one in terms of evaluation activity, but the number of comments is quite different.

The last cluster (red) is the smallest one. It consists of four absolute project leaders, who together with some of the yellow participants turned out to be winners or winning ideas authors.

For greater classification veracity the obtained clusters were modified: some of the grey, blue and green balls formed a new class of creators, the rest of the blue joined the grey cluster; also, some minor rearrangements were made.

## 6    Results

Table 2 represents the resulting user types, their percentages, descriptions and equivalents in other studies.

**Table 2.** Types of collaboration platform Witology participants

| User type | Number / % of objects | | Description | User types of previous studies |
|---|---|---|---|---|
| Celebrities | 4 | 1% | Outstanding users, champions. | Actives [5], mavens [13], active participants [7], insiders [23], celebrities [24] |
| Debators | 21 | 4% | Those who comment and evaluate actively. | Debators/socializers [5], connectors/salesmen [13], active participants [7], minglers [23] |
| Creators | 20 | 4% | Idea generators. Could be divided into two groups: energetic creators (6 users), who not only create, and sociopathic ones (14 users), | Mavens [13], creators [25], active participants/sporadic contributors [7], insiders/devotees [23] |

| | | | who comment or evaluate many times less. | |
|---|---|---|---|---|
| Critics | 34 | 7% | Those who evaluate but don't meddle in discussions. | Critics/spectators [25], sporadic contributors [7], lurkers [24] |
| Tourists | 177 | 35% | Those who rarely make attempts to participate. | Sporadics/lurkers [5], spectators [25], lurkers [7], tourists [23], newbies/lurkers [24] |
| Inactives | 248 | 49% | Those who do absolutely nothing. | Sporadics/lurkers [5], inactives[25], lurkers [7], tourists [23] |

The developed typology and type percentages can be compared with two rather general typologies from the top of the lattice (fig. 1). Table 3 shows how the six classes of this research correspond to their classes.

**Table 3.** Comparison of different typologies class percentages

| Nielsen | % | Brandtzæg | % | This study | % |
|---|---|---|---|---|---|
| Active participants | 1% | Actives | 18% | Celebrity Debators | 5% |
| Sporadic contributors | 9% | Debators Socializers | 36% | Creators Critics | 11% |
| Lurkers | 90% | Lurkers Sporadics | 46% | Tourists Inactives | 84% |

Interestingly, the percentages in the obtained typology are very close to the ones in Nielsen typology. Brandtzæg explains the discrepancy with the "90 – 9 – 1" rule by a relatively low popularity of Norway social networks compared to YouTube or Wikipedia and by smaller content creation barriers, but such an explanation is not likely to be relevant for the given collaboration project. Nearly 90% of lurkers could be accounted for by initially a small interest of participants to the work itself and a great curiosity to a new for Russia phenomenon, crowdsourcing, as means of some company's growth and development. Other reasons may also take place, but it seems to be difficult to identify them without several projects or platforms comparison.

## 7    Conclusions

During the process of literature exploration it appeared that there is no generally accepted SNS user classification or any specific collaboration platform participant ty-pology. Based on the existing relevant typologies of social networks, blogs, news-groups users by means of cluster analysis we developed an original collaboration platform typology. The six classes are so far not expected to be suitable for other crowdsourcing communities. The percentages of classes follow the rule "90 − 9 − 1", according to which only a minor part of the community is really active.

Thus, all the research objectives were mainly attained.

### 7.1    Future Work

The developed typology is far from being complete and final. Only a small sample of one of the project was analyzed, while different projects data comparison is expected to specify the classification greatly. Possible future work also includes the following:

- Involving more diverse information on the project (e.g. logs, qualitative values of user evaluations).
- Using other methods (factor analysis, graph mining, mean analysis) of group detection or other clustering algorithms.
- Finding special users (e. g. trolls, flamers, flooders [24]).
- Developing a classification algorithm.
- Testing connection between group membership and demographical factors (age, sex) or psychological tests results.
- Using special metrics to determine community health [14].

Judging by the number of possible work improvement directions it can be concluded that this paper is only a small test sally into the investigation of collaboration platform participants' behavior, which describes only a static snapshot of one project and does not claim to be indisputable and fundamental.

**Acknowledgements.** This work was partially done during the mutual research project between Witology and Higher School of Economics (Project and studying group "Data Mining algorithms for analysing Web forums of innovation projects discussion"). We would like to thank Jonas Poelmans for his suggestions for improving the paper.